\documentclass[prl,twocolumn,superscriptaddress,showpacs,floats,floatfix]{revtex4}
\usepackage{epsfig}

\begin{document}

\preprint{1}

\title{Evidence for a single hydrogen molecule connected
by an atomic chain}

\author{M. Kiguchi}
\thanks{Present address: Department of Chemistry, Hokkaido
University, Sapporo 060-0810, Japan.} \affiliation{Kamerlingh
Onnes Laboratorium, Universiteit Leiden, PO Box 9504, NL-2300 RA
Leiden, The Netherlands}

\author{R. Stadler}
\author{I. S. Kristensen}
\affiliation{Center for Atomic-scale Materials Physics, Department
of Physics, NanoDTU, Technical University of Denmark, DK-2800 Kgs.
Lyngby, Denmark}

\author{D. Djukic}
\affiliation{Kamerlingh Onnes Laboratorium, Universiteit Leiden,
PO Box 9504, NL-2300 RA Leiden, The Netherlands}

\author{J.\,M.\ van Ruitenbeek}
\thanks{Electronic mail address: ruitenbeek@physics.leidenuniv.nl}
\affiliation{Kamerlingh Onnes Laboratorium, Universiteit Leiden,
PO Box 9504, NL-2300 RA Leiden, The Netherlands}

\date{\today}

\begin{abstract}
Stable, single-molecule conducting-bridge configurations are
typically identified from peak structures in a conductance
histogram. In previous work on Pt with H$_2$  at cryogenic
temperatures it has been shown that a peak near 1 $G{_0}$
identifies a single molecule Pt-H$_{2}$-Pt bridge. The histogram
shows an additional structure with lower conductance that has not
been identified. Here, we show that it is likely due to a hydrogen
decorated Pt chain in contact with the H$_2$ molecular bridge.
\end{abstract}

\pacs{73.63.Rt, 63.22.+m, 73.23.-b, 85.65.+h}

\maketitle

The interest in chains of single metal atoms bridging between two
electrodes is largely due to their unique properties as ideal
one-dimensional systems \cite{1}. For clean metals, only Au, Pt
and Ir form atomic chains \cite{3,4}. However, atomic or molecular
adsorption on metal surfaces can widen this scope. Recently, 2 nm
long Ag atomic chains have been created in the presence of oxygen
at ultra low temperature, while clean Ag only forms short chains
\cite{5}. Atomic chains have been imaged by transmission electron
microscope(TEM) for the noble metals Cu, Ag, and Au
\cite{23,24,25,26}.

These atomic wires could form ideal leads to organic molecules for
the investigation of their potential use in molecular electronics
\cite{11}. The first experiments contacting molecules by Pt atomic
leads appear to indicate that atomic chain formation is suppressed
by the molecular adsorption \cite{9,10}. In the latter works it
was shown that a single hydrogen molecule, H$_{2}$, can be
contacted between Pt leads. The hydrogen molecule is a valuable
model system for single-molecule junctions. By use of point
contact spectroscopy, isotope substitution and shot noise
measurements the system was characterized in great detail and
close agreement with atomistic model calculations was obtained
\cite{9,10}. The Pt-H$_{2}$-Pt junction was first identified by
its conductance. It shows up as a recurring plateau in the
conductance when controllably breaking a contact, and in a
histogram of conductance values collected for many such breakings
it gives rise to a sharp peak near 1 $G_{0}$, where 1
$G_{0}$=$2e^{2}/h$ is the conductance quantum. This main peak at 1
$G_{0}$ for the Pt/H$_{2}$ system is therefore well understood.
However, there is more structure in the conductance histogram for
the Pt/H$_{2}$ junctions, which has not been explained. In
particular, a strong peak is found at about 0.1-0.2 $G_{0}$,
suggesting that other configurations of hydrogen between Pt leads
may be formed. In the present study we focus on those structures,
having a conductance below 1 $G_{0}$, and we present evidence that
they can be attributed to the formation of a hydrogen decorated Pt
atomic chain, that forms one of the leads contacting a hydrogen
molecule.

The measurements have been performed using the mechanically
controllable break junction technique (see Ref.~\onlinecite{27}
for a detailed description). Once under vacuum and cooled to 4.2 K
a fine Pt wire was broken. Atomic-sized contacts between the wire
ends can be formed using a piezo element for fine adjustment.
H$_{2}$ was admitted via a capillary. DC two-point voltage-biased
conductance measurements were performed by applying a voltage in
the range from 10 to 150 mV. Every statistical data set (such as a
conductance histogram) was build from a large number (over 3000)
of individual digitized conductance traces. AC voltage bias
conductance measurements were performed using a standard lock-in
technique. The conductance was recorded for fixed contact
configuration using an ac modulation of 1 mV amplitude and a
frequency of 7.777 kHz, while slowly ramping the DC bias between
-100 and + 100 mV.

\begin{figure}[!b]
\epsfig{width=7.8cm,figure=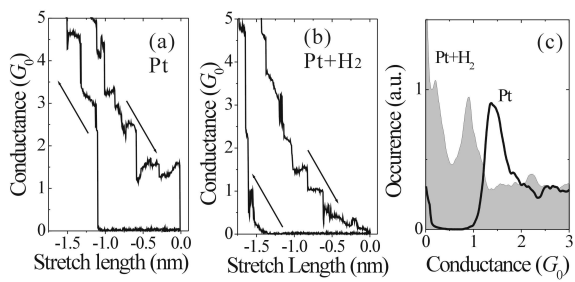} \caption{ Breaking and return
traces for clean Pt (a), and for Pt in a H$_{2}$ atmosphere (b).
Many of such curves are collected into conductance histograms as
shown in (c) for clean Pt (black curve) and Pt in H$_{2}$ (filled
graph).} \label{fig1}
\end{figure}

Figures~\ref{fig1} (a) and (b) show typical conductance traces for
clean Pt and for Pt after admitting H$_{2}$. After admitting
H$_{2}$, plateaux near 1$G_{0}$ are frequently observed and the
corresponding histogram (Fig.~\ref{fig1}(c)) shows a sharp feature
near 1 $G_{0}$. The plateaux near 1$G_{0}$ and the corresponding
peak in the histogram originate from single-molecule Pt-H$_{2}$-Pt
contacts, as shown by previous studies \cite{9,10}. In addition to
the 1$G_{0}$ feature, the histogram shows a peak near 0.2 $G_{0}$
on top of a low-conductance tail. Looking at the individual
traces, we find that the conductance does not suddenly drop, as
expected for a contact break after the appearance of the plateau
near 1$G_{0}$, but rather decreases by small steps while the
contact is being stretched. This suggests that a structure having
a conductance below 1$G_{0}$ is formed by further stretching the
single-molecule Pt-H$_{2}$-Pt contact. The conductance trace in
Fig.~\ref{fig1}(b) shows that the structure can be stretched to
quite long lengths ($>$0.5 nm in Fig.~\ref{fig1}(b)), which
suggests the formation of an atomic chain.

\begin{figure}[!t]
\epsfig{width=8cm, figure=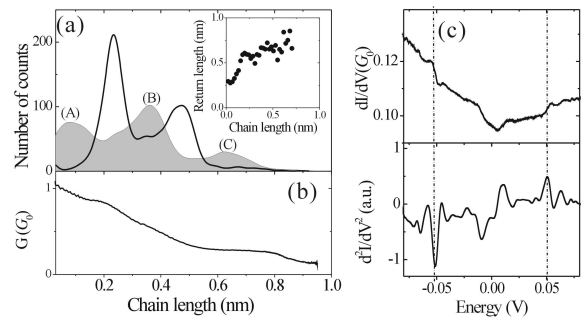}
\caption{ (a) Length histogram
for clean Pt (black curve) and Pt in H$_{2}$ (filled graph). The
start and stop values between which the lengths were measured were
taken as (2.5, 1.0) for Pt and (1.3, 0.1) for Pt/H$_{2}$, in units
of $G_{0}$. Inset: Average return lengths as a function of chain
length. (b) Average conductance as a function of chain length for
Pt/H$_{2}$. (c) Differential conductance (top) and its derivative
(bottom) for a Pt/ H$_{2}$ contact taken at a conductance of ~0.1
$G_{0}$.} \label{fig2}
\end{figure}

In order to investigate the chain formation, we measure the length
histogram of the last plateau and the return length distribution.
Figure~\ref{fig2}(a) shows the length histogram for the final
conductance plateaux for the Pt/ H$_{2}$ contacts (filled grey
distribution), and this is compared to a length histogram for
clean Pt. The length for Pt/H$_{2}$ is taken here as the distance
between the points at which the conductance drops below 1.3
$G_{0}$ and 0.1 $G_{0}$, respectively, while for clean Pt the
boundaries are 2.5 $G_{0}$ and 1.0 $G_{0}$. The former boundaries
are set such as to capture the length of the 1$G_{0}$-plateau plus
the subsequent structures that give rise to the peak around 0.2
$G_{0}$ in the conductance histogram. It is striking that the
Pt/H$_{2}$ contact can be stretched as long as 0.8 nm. A sequence
of peaks is observed in the Pt/H$_{2}$ length histogram of
Fig.~\ref{fig2}(a), indicating the repeated occurrence of certain
stable chain configurations that we identify as (A), (B), and (C).
The distance between the peaks is 0.27$\pm$0.01 nm, which is
slightly larger than the Pt-Pt distance of a clean Pt atomic chain
(0.23 nm) \cite{4}.

The inset of Fig.~\ref{fig2} (a) shows the average return lengths
as a function of chain length. This is the distance over which the
two electrodes need to be moved back after the junction breaks in
order to re-establish contact, averaged over many break cycles.
Apart from an offset of 0.3 nm due to the elastic response of the
banks \cite{4}, the relation is approximately proportional,
suggesting that a fragile structure is formed with a length
corresponding to that of the last plateau, which is unable to
support itself when it breaks and collapses onto the banks on
either side.

We further test this interpretation by analyzing the stretch
length dependence of the conductance and by point contact
spectroscopy. Figure~\ref{fig2}(b) shows the average conductance
for Pt/H$_{2}$ junctions as a function of the chain length. The
curve is obtained by adding all measured conductance traces from
the start value (1.3 $G_{0}$) onward, and dividing at each length
by the number of traces included at that point. The mean
conductance decreases rapidly as the chain becomes longer.
Although the conductance for a pure Pt chain also decreases with
length \cite{28} its conductance stays well above 1 $G_{0}$. This
fact, combined with the larger peak distance in the length
histogram indicate that the structure with lower conductance is
probably not a clean Pt atomic chain, and may be due to a hydrogen
decorated atomic chain. The average conductances of structures
(A), (B), and (C) are 0.96, 0.56 and 0.28 $G_{0}$, respectively.
Since the conductance of structure (A) is close to 1 $G_{0}$, we
identify it with the single molecule Pt-H$_{2}$-Pt contacts that
have been studied previously \cite{9,10}. We discuss the new
structures (B) and (C) that arise by further stretching of the
Pt-H$_{2}$-Pt junction in the following. Note that a stable level
near 0.2-0.3 $G_{0}$ is observed at 0.6-0.8 nm in length in
Fig.~\ref{fig2}(b). The slow length dependence gives rise to a
high number of counts in a conductance histogram, which explains
the peak at 0.2 $G_{0}$ in Fig.~\ref{fig1}(c). Apart from this
stable structure that we have labelled (C) the length histogram
points at an intermediate structure (B), for which the conductance
varies more strongly with stretching.

Figure~\ref{fig2}(c) shows  an example of the differential
conductance and its derivative for a Pt/H$_{2}$ contact taken at a
conductance of 0.1 $G_{0}$. The spectrum shows an increase in the
differential conductance symmetrically at 51 meV, and clear
symmetric peaks are observed in the second derivative,
$d^{2}I/dV^{2}$. The symmetric peaks are commonly observed near
$\sim$57$\pm$4 meV in the second derivatives for contacts having
conductances in the range 0.6-0.1 $G_{0}$. Note that the sign of
the signal in Fig.~\ref{fig2}(c) agrees with that for inelastic
electron tunnelling spectroscopy, as expected for low conductance
\cite{smit}. However, most signals that we find are of the type
discussed in Ref.~\onlinecite{29}. The energy of 57 meV agrees
with the energy of the transverse translation mode of the molecule
in the Pt-H-H-Pt configuration \cite{10}. The close agreement
suggest that a hydrogen molecule is still bridging the junction
after stretching it beyond the 1 $G_{0}$ plateau.

Based on the experiments presented above we arrive at the
following chain formation model for Pt atomic contacts with
H$_{2}$. First, a single hydrogen molecule is adsorbed between Pt
electrodes (structure A). Further stretching induces the
incorporation of the first Pt atom from the stem part of the
electrode into the chain (structure B). Then, the second Pt atom
is incorporated into the chain (structure C), and the atomic chain
is formed with a single hydrogen molecule bridging. The larger
distances in the length histogram and the low conductance should
then be attributed to additional hydrogen decorating the Pt atomic
chain. Further support for this chain formation process is
obtained from density functional theory (DFT) calculations.

Electronic structure calculations were performed using a plane
wave implementation of DFT \cite{dacapo} with an energy cutoff of
340 eV, where we used ultra-soft pseudopotentials
\cite{vanderbilt90}, and a PW91 parametrization for the exchange
and correlation functional \cite{pw91}. The transmission functions
of the molecular junctions were calculated using a general
non-equilibrium Green's function formalism for phase-coherent
electron transport \cite{meir92}, where both, the Green's function
of the scattering region and the self-energies describing the
coupling to the semi-infinite electrodes, were evaluated in terms
of a basis consisting of maximally localized Wannier functions
\cite{method}. In our calculations the supercells for the
scattering region are defined by 3$\times$3 atoms in the surface
plane and contain three to four surface layers on each side of the
molecule. We used a 4$\times$4 grid for the k-point integration in
order to obtain well converged results for the conductance
\cite{kpoints}.

\begin{figure}[!t]
\epsfig{ width=7cm, figure=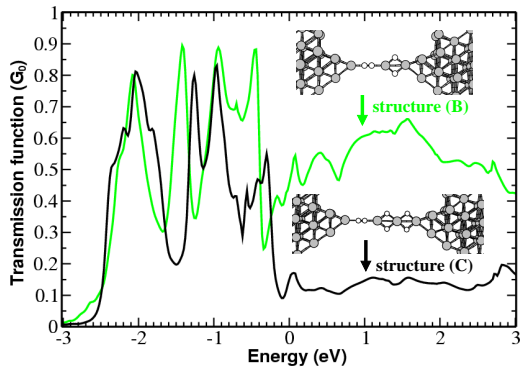} \caption{(Color online)
Transmission functions for the proposed structures (B) (green/gray
curve) and (C) (black curve), respectively, as calculated from DFT
for a setup with bulk electrodes and a (111) surface with a
pyramid of Pt atoms on top. The zero point of the x-axis is the
Fermi energy. The explicit geometries of the structures are shown
as insets.} \label{fig3}
\end{figure}

Based upon the distances and conductances found in the experiment
presented above we came to consider the model structures for (B)
and (C) illustrated by the insets in Fig.~\ref{fig3}. The figure
shows calculated transmission functions for structures (B) and (C)
as a function of energy. The transmission function for structure
(A), a hydrogen bridge with no additional hydrogen having a
conductance of 1 $G_{0}$ \cite{prl94} is lowered when going from
(A) to (C). The conductances are found as 0.46 and 0.15 $G_{0}$
for structures (B) and (C), respectively, which is in reasonable
agreement with the experimental values of $\sim$0.6 and $\sim$0.3
$G_{0}$. The theoretical values have been obtained after
optimizing the distance between the contacts by total energy
minimization and the optimal length of a Pt wire segment decorated
with additional hydrogen was found to be 0.272 nm which agrees
with the experimental results.

\begin{figure}[!t]
\epsfig{ width=55mm, figure=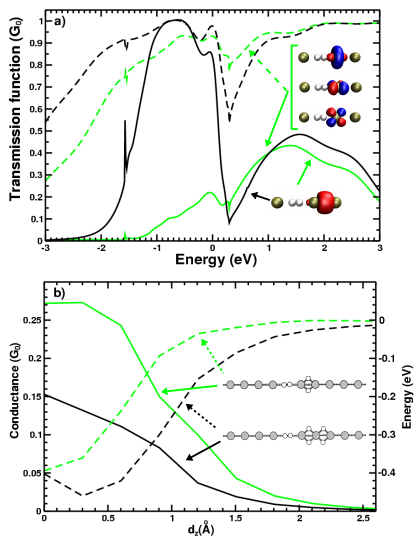}
\caption{(Color online) (a)
Transmission functions obtained from calculations of a single
H$_{2}$ bridge between wires of Pt atoms with the s-orbital (solid
black curve), the d-orbitals (dashed green/grey curve), both the
s- and d-orbitals (solid green/grey curve) and no atomic orbitals
(dashed black line) removed from the Hamiltonian. The removed
orbitals are also shown superimposed with the atomic positions in
the insets. (b) Conductance (solid curves) and binding energy
(dashed curves) for the molecule inside the junction for
structures similar to (B) (green/grey) and (C) (black) assumed to
have linear atomic wire arrangements as shown in the insets.}
\label{fig4}
\end{figure}

Our analysis shows that this reduction is due to the additional
hydrogen atoms saturating the s-orbital and part of the d-orbitals
between the Pt atoms they are attached to, thereby making them
unavailable for electron transport. This is illustrated in
Fig.~\ref{fig4}(a) that shows the results of calculations for the
H$_{2}$ bridge structure without additional hydrogen. We have
simulated the effect hydrogen addition might have by cutting out
Pt Wannier functions from the scattering Hamiltonian. The removal
of the Pt s-orbital for one and two subsequent wire atoms (shown
only for the first case in Fig.~\ref{fig4}(a)) accounts for the
successive reduction of the transmission at around 1 eV and higher
above the Fermi level. For the reduction of the transmission
directly at $E_{F}$ and therefore the conductance, also a blocking
of some of the d-orbitals is needed, which is also shown in
Fig.~\ref{fig4}(a). Although a cutting of the d-Wannier functions
of the first Pt-atom alone does not have a large effect on the
conductance and seems to even enhance it, the conductance is
drastically reduced if both the s- and d-orbitals of the same Pt
atom are blocked. For structure (B) we also compared the
transmission functions of the atomic wire system with that of the
more realistic surface calculation in Fig.~\ref{fig3}.
Qualitatively, the results are similar in terms of the main peak
structure illustrating that the analysis is robust with respect to
details of the atomic arrangement. In a quantitative comparison,
however, the conductance from the surface calculation is higher
than for the wire system by approximately a factor 2.0.

In Fig.~\ref{fig4}(b) we show the dependence of the conductance
with increasing the distance between the wire electrodes. In these
calculations the positions of all hydrogen atoms and four Pt atoms
on each side have been fully relaxed. The corresponding binding
energy of the H$_{2}$ bridge molecule to the wires is also shown
for comparison. Although there is a short plateau in the
conductance for structure (B) when stretched  beyond the optimal
bonding distance (green/grey curves in Fig.~\ref{fig4}(b)), for
structure C (black curves) the overall decrease with the length
seems to be slower up to a stretching length of 0.1 nm. This trend
was also found in the experimental histogram. The shift in optimal
distance between (B) and (C) just reflects the increase of the
total cell length due to the expansion of a second wire segment.

Let us comment on the path that may lead to the structures
discussed above. Since the bridging H$_{2}$ molecule is very
weakly bound to the Pt electrodes, 'wire pulling' seems to be
unlikely.  Structures (B) and (C) can only be formed by a
concerted process. Such a process may involve the displacement of
Pt atoms at the surface due to phonons and the formation of
intermediate structures with the additional hydrogen adsorbed on
the electrodes. The configuration space for covering all
possibilities for atomic movements in such a process is too large
to be explored by our calculations in detail.

In conclusion, the experimental evidence shows that a
Pt-H$_{2}$-Pt single-molecule junction can be stretched further
into forming an atomic wire. We propose a likely structure for
this wire in terms of a Pt atomic chain decorated with hydrogen.
This interpretation is supported by DFT calculations. While the
pathway that brings new atoms into the atomic chain structure
remains problematic, we obtain fair agreement in the numbers for
the bond distances and the conductances.

We are very grateful to K. W. Jacobsen for his support and for
many stimulating discussions. This work is part of the research
program of the Stichting FOM, which is financially supported by
NWO. The Center for Atomic-scale Materials Physics at NanoDTU is
sponsored by the Danish National Research Foundation. We
acknowledge support from the Nano-Science Center at the University
of Copenhagen and from the Danish Center for Scientific Computing
through Grant No. HDW-1101-05.

\end{document}